\documentclass[12pt,noshowpacs,nofootinbib,notitlepage,amsmath]{revtex4-1}
\usepackage{setspace}
\allowdisplaybreaks
\usepackage{graphicx,color}
\usepackage[colorlinks=true,citecolor=blue,linkcolor=blue,urlcolor=blue]{hyperref}
\usepackage[charter]{mathdesign}
\usepackage{multirow}
\usepackage{enumerate}

\DeclareSymbolFontAlphabet{\mathcal}{symbols}
\DeclareSymbolFont{symbols}{OMS}{xmdcmsy}{m}{n}
\DeclareSymbolFont{largesymbols}{OMX}{xmdcmex}{m}{n}
\SetSymbolFont{symbols}{bold}{OMS}{xmdcmsy}{b}{n}

\begin{document}  
\title{\color{blue}\Large Damping of gravitational waves in 2-2-holes}

\author{Bob Holdom}
\email{bob.holdom@utoronto.ca}
\affiliation{Department of Physics, University of Toronto, Toronto, Ontario, Canada  M5S 1A7}
\begin{abstract}
A 2-2-hole is an explicit realization of a horizonless object that can still very closely resemble a BH. An ordinary relativistic gas can serve as the matter source for the 2-2-hole solution of quadratic gravity, and this leads to a calculable area-law entropy. Here we show that it also leads to an estimate of the damping of a gravitational wave as it travels to the center of the 2-2-hole and back out again. We identify two frequency dependent effects that greatly diminish the damping. Spinning 2-2-hole solutions are not known, but we are still able to consider some spin dependent effects. The frequency and spin dependence of the damping helps to determine the possible echo resonance signal from the rotating remnants of merger events. It also controls the fate of the ergoregion instability.
\end{abstract}
\maketitle

\section{Introduction}
A 2-2-hole \cite{Holdom:2016nek} is an example of an extremely compact horizonless object that deviates from a black hole (BH) only within some Planck-like distance from the would-be horizon. 2-2-holes exist as a class of solutions to the field equations of quadratic gravity whose only mass scale is the Planck mass. Thus the extreme compactness, as characterized by the Planck-like distance, is of purely gravitational origin. At the same time there is no upper limit on the size of these solutions, and given that they are nearly indistinguishable from BHs in many ways, there is the question of whether all BHs are actually 2-2-holes. Here we consider 2-2-holes that are gravitationally bound balls of a relativistic gas, solutions that were found and studied in \cite{Holdom:2019ouz} and \cite{Ren:2019jft}, when the gravitational action consists of the Einstein term $R$ and the Weyl-tensor-squared term $C_{\mu\nu\alpha\beta}C^{\mu\nu\alpha\beta}$ only. The Weyl term is essential for the existence of 2-2-hole solutions, and the solutions are expected to be qualitatively similar when a $R^2$ term is added.

These solutions are described in terms of the line element
\begin{align}
ds^2=-B(r)^2 dt^2 + A(r)^2 dr^2 +r^2 d\Omega^2
\end{align}
where $A(r)=a_2 r^2+\dots$ and $B(r)=b_2 r^2+\dots$ near the origin. The relativistic gas with $\rho(r)=3p(r)$ has temperature $T(r)=T_\infty B(r)^{-1/2}$. The field equations determine $A(r), B(r)$ and the combination $N T_\infty^4 m_2^2/m_\textrm{Pl}^2$, where $N$ is $g_b+7g_f/8$ with $g_b$ and $g_f$ being the number of bosonic and fermionic degrees of freedom in nature, and $m_2$ is the mass of the unstable spin-2 mode of quadratic gravity. The cumbersome field equations must be dealt with numerically, but this is sufficient to not only find these solutions but also to find the scaling behavior that relates 2-2-holes of different sizes. With the solution in hand the entropy is explicitly calculable,
\begin{align}
S=\frac{(2\pi)^3}{45}N\int_0^{3GM} T(r)^3A(r)^{1/2}r^2dr=\zeta\frac{\textrm{Area}}{4G}=\zeta S_{\rm BH}\label{e9}
\end{align}
where $\zeta\approx0.75 N^\frac{1}{4}\sqrt{m_2/m_{\rm Pl}}$. The integrand is finite and it peaks at $r=0$. The metric functions in the interior of the 2-2-hole have a strong scaling dependence on $M$ (e.g.~$a_2,b_2\sim G/(GM)^4$) and this is responsible for the area-law scaling for $S$. The 2-2-hole also satisfies $ST_\infty=S_{\rm BH}T_{\rm Hawking}$ and so $T_\infty=T_{\rm Hawking}/\zeta$. The total matter energy is $U=\frac{3}{8}M$.

A 2-2-hole is distinguishable from a BH in at least one way, and that is in the behavior of low frequency gravitational waves in its vicinity. The description of a gravitational wave around a 2-2-hole reduces to a radial equation that describes a wave in a 1D cavity. Low frequencies waves tend to be trapped due to a boundary condition at one end ($r=0$) and a potential barrier at the other end ($r\approx3GM$). The cavity structure of a 2-2-hole implies a resonance spectrum \cite{Conklin:2017lwb} and possible echoes in the time domain \cite{Cardoso:2016rao,Cardoso:2016oxy,Abedi:2016hgu}. But how is this picture influenced by the gravitational wave interacting with the matter inside the 2-2-hole? The properties of the gas are well determined as we have just described, and this enables this question to be addressed.

The wave and cavity description is easiest when using the tortoise radial coordinate $x$ defined such that the line element is
\begin{align}
ds^2=B(r)^2 (-dt^2 + dx^2) +r^2 d\Omega^2.
\end{align}
In this coordinate system waves travelling in the radial direction have unit speed and they propagate as planes waves away from the potential barrier. We can let $x=0$ correspond to $r=0$ and let $\Delta x$ be the size of the cavity. The round trip travel time is $\Delta t=2\Delta x=2GML$ where $L$ is the dimensionless length of the cavity,
\begin{align}
   L=2\,\eta\log(\frac{GM}{\ell_{\rm Pl}})
.\label{e8}\end{align}
The cavity may be said to be stretched due to the large log associated with the Planck-like distance mentioned at the beginning. The stretching of the tortoise coordinate occurs for $r$ where $A(r)/B(r)$ is large, and this ratio reaches a peak value at the Planck-like distance from the would-be horizon. In (\ref{e8}) this distance is encoded in the parameter $\eta$; if $\eta=1$ then this distance is the Planck length $\ell_{\rm Pl}$ and if $\eta=2$ it is the much smaller proper Planck length. $\eta$ describes a feature of the 2-2-hole solution that does not scale in a simple way, and numerical determination of $\eta$ for large 2-2-holes is difficult. If the resonance spectrum is observed then $\eta$ will be empirically determined, as suggested in \cite{Holdom:2019bdv}. $L=400$ is a representative value.

\begin{figure}[h]
\centering
\includegraphics[width=0.65\textwidth]{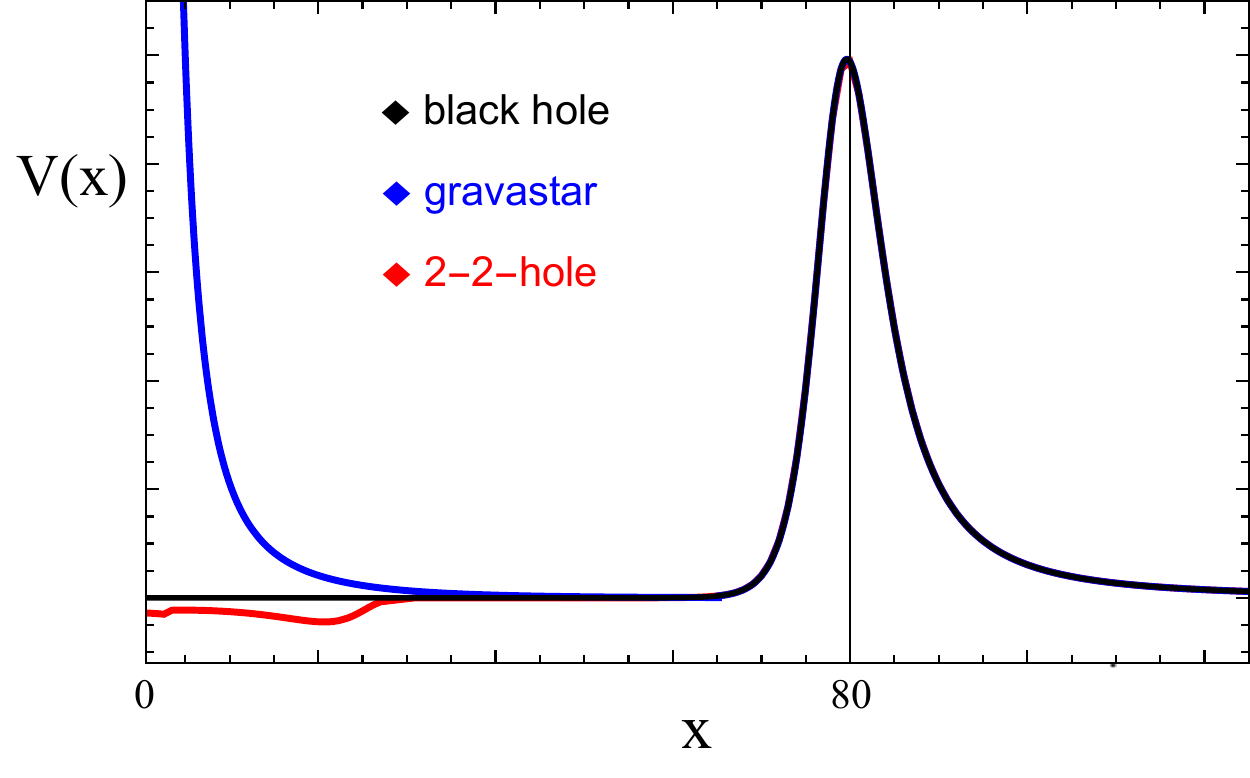}  
\caption{The potential in the scalar wave equation and the resulting cavity structure is compared for three compact objects, pictured for $L=80$ with $GM=1$ (reproduced from \cite{Holdom:2016nek}). The origin of the 2-2-hole, the origin of the gravastar and the radius at which the BH geometry is truncated for the truncated BH construction, all correspond to the tortoise coordinate $x=0$ on this plot. The dip in the potential for the 2-2-hole corresponds to the 2-2-hole interior, and the length of this dip remains the same for a larger more realistic $L$.}
\label{fig00}
\end{figure}

In Fig.~\ref{fig00} we compare the effective potentials appearing in the massless scalar field wave equation for three different backgrounds: a 2-2-hole, a very compact gravastar and a truncated BH. The cavity size shown is $L=80$. The 2-2-hole lacks the divergent angular momentum barrier at $r=0$ that would be present for a flat or any nonsingular geometry, such as the gravastar. This finiteness of the 2-2-hole potential is related to the existence of a timelike curvature singularity at $r=0$, and thus the diverging curvatures in the interior of a 2-2-hole have a benign effect on the wave equation. The effect is similar to introducing a wall just slightly outside the horizon of a BH, which is the truncated BH construction. The interior of the 2-2-hole corresponds to the small region in the cavity where the 2-2-hole potential deviates from the truncated BH; this region is relatively even smaller for $L=400$ rather than $L=80$. 

The propagation of gravitational waves around a 2-2-hole has the standard description from the Einstein equations both inside and outside the cavity except for this small region near the end of the cavity, corresponding to the 2-2-hole interior. The gravitational field equations in this high curvature region would need to be obtained from the quadratic gravity action. This region also contains the matter, the relativistic gas, of interest here for its possible damping effect. The large number of particle species, all of which are excited in the interior, is much larger than the number of gravitational degrees of freedom. The gravitational interactions in quadratic gravity are also asymptotically free in the high curvature limit. For these reasons we expect that the matter gives the leading effect for the question of damping.

The thermodynamic properties of the gas is determined by the 2-2-hole solution, as we have outlined, and how this determines the damping is the focus of section \ref{s2}. In section \ref{s3} we extend our results to rotating objects. In this case the 2-2-hole solutions are not known and so we lack the precise thermodynamical description, although we expect the qualitative features to be similar. What we can more precisely account for is the effects of spin on the propagation of gravitational waves exterior to a rotating 2-2-hole. We thus study the spin dependent damping in the context of the truncated Kerr BH model, which is a popular model for rotating objects of extreme compactness. It shares with the truncated Schwarzchild model a non-diverging potential, a property we also expect for the rotating 2-2-hole. Gravitational perturbations on a truncated Kerr BH background leads to the ergoregion instability \cite{fried,Brito:2015oca}, and so we are then able to estimate the damping needed to avoid this instability. The role that damping can have on the ergoregion instability of extremely compact objects has been studied previously in \cite{Maggio:2017ivp,Maggio:2018ivz}.

\section{Damping in the cavity}\label{s2}
To study damping we start with the Hawking formula \cite{Hawking:1966qi} for the rate of energy loss experienced by a gravitational wave traveling through a medium with a shear viscosity. We choose to write this viscosity as the viscosity-to-entropy density ratio $\cal V$ times the entropy density $s$,
\begin{align}
\frac{1}{E}\frac{dE}{dt}&=-16\pi G{\cal V}s
\label{e1}.\end{align}
This simple result needs to be corrected in several ways for our study. A serious deficiency of (\ref{e1}) is that it only applies when the frequency of the wave is much less than the collisional frequency in the medium. This is typically not the case and we shall describe a correction factor below. Secondly, we shall also account for an effect that occurs when the wavelength becomes larger than the size of the matter region. Thirdly, (\ref{e1}) assumes a flat space form for the wave equation. This may not be a bad approximation for the radial scalar field wave equation in the interior of a 2-2-hole, given the small potential in that region in Fig.~\ref{fig00}. Whatever the analogous result is for the gravitational wave equation, we expect its effect on the damping result to be sub-dominant to the previous two effects.

The collisional frequency $\nu_c$ is proportional to the scattering cross section between particles which in turn is proportional to the square of some effective coupling strength. We are considering a highly relativistic gas and so we can write
\begin{align}
   \nu_c(x)=\bar{\alpha}^2 T(x)
,\label{e2}\end{align}
where $\bar{\alpha}$ is the dimensionless coupling and $T$ is the temperature. We shall use this relation to define $\bar{\alpha}$. The viscosity-to-entropy-density ratio ${\cal V}(\bar{\alpha})$ will depend on this coupling. Perturbative QCD for instance gives ${\cal V}(\bar{\alpha})\propto 1/(\bar{\alpha}^2\ln\bar{\alpha}^{-1})$ \cite{Arnold:2000dr}. We might not be in the regime of small coupling and so we keep the general function in the following.

We are assuming that the system can be characterized by a single $\bar{\alpha}$. Thus we are ignoring the dependence of the coupling on the spatially varying temperature due to the energy dependence of running couplings. We also ignore the different couplings through which the different particle species in the gas can interact. It remains to refine the present work to include these effects.

The matter in the cavity has an entropy density $s(x)$, the entropy per unit area per unit $x$. The time $t$ in (\ref{e1}) may be traded for $x$ and we further trade $x$ for a dimensionless coordinate $\tilde{x}=x/(GM)$ and define $\tilde{s}(\tilde{x})=GM\,s(GM\tilde{x})$ so that
\begin{align}
\frac{1}{E}\frac{dE}{d\tilde{x}}&=-16\pi{\cal V}(\bar{\alpha}) G\,\tilde{s}(\tilde{x})
.\end{align}
From this we can define a damping factor $R_{\rm damp}$ that gives the fraction of wave energy remaining after the round trip travel time $\Delta t=2GML$,
\begin{align}
R_{\rm damp}=\exp\left(-32\pi{\cal V}(\bar{\alpha}) G\int_0^L\tilde{s}(\tilde{x})d\tilde{x}\right)
.\label{e7}\end{align}
$R_{\rm damp}=1$ means no damping.

The integral factor in (\ref{e7}) is the entropy per unit area and so from the result in (\ref{e9}) we have
\begin{align}
\int_0^L\tilde{s}(\tilde{x})d\tilde{x}=\int_0^{\Delta x} s(x)dx=\frac{S}{\rm Area}=\frac{\zeta}{4G}
.\label{e5}\end{align}
(\ref{e5}) then gives a simple result for the damping factor,
\begin{align}
R_{\rm damp}=\exp\left(-8\pi{\cal V}(\bar{\alpha})\zeta\right)
.\label{e4}\end{align}
Both $G$ and $M$ have dropped out, as made possible by having a self-gravitating system. For a typical value of $\zeta$ we use 2.5, for example when $N\approx 125$ and $m_2\approx m_{\rm Pl}$.

Since ${\cal V}(\bar{\alpha})$ has a theoretical lower bound of $1/(4\pi)$ \cite{Kovtun:2004de}, the result in (\ref{e4}) implies strong damping independent of the gravitational wave frequency $\omega$. But this is not correct. To account for an $\omega$ that is not much less than the collisional frequency $\nu_c$, a frequency dependent suppression factor $(1+(\omega/\nu_c)^2)^{-1}$ should be introduced to modify the Hawking damping rate (\ref{e1}). This factor has been discussed recently in \cite{Loeb:2020lwa}. It is present because the shear induced by the gravitational wave changes sign over the period of the wave, and so the time over which the particle can sample the shear is limited by this time, rather than the collisional time, when $\omega$ is larger than $\nu_c$. For a given $\nu_c$ and when $\omega\gg\nu_c$, the damping rate is reduced by a factor $\nu_c^2/\omega^2$. As for the $\nu_c$ dependence, the viscosity typically already has a $1/\nu_c$ dependence, and so the damping rate increases linearly with $\nu_c$ up to $\nu_c\approx\omega$, after which it falls like $1/\nu_c$.

For our problem we should consider the proper frequency $\omega(x)=\omega B(r(x))^{-1/2}$, and then the required ratio is $\omega(x)/\nu_c(x)=\omega/(\bar{\alpha}^2T_\infty)$ from (\ref{e2}) with $T(x)=T_\infty B(r(x))^{-1/2}$. And we have already noted that $T_\infty=T_{\rm Hawking}/\zeta=1/(8\pi GM\zeta)$. In terms of the dimensionless frequency $\tilde{\omega}=GM\omega$ we now have a frequency dependent damping factor,
\begin{align}
R_{\rm damp}(\tilde{\omega})=\exp\left(-8\pi{\cal V}(\bar{\alpha})\zeta\left[1+\left(8\pi\tilde{\omega}\frac{\zeta}{\bar{\alpha}^2}\right)^2\right]^{-1}\right)
.\end{align}
This additional factor in the exponent quite dramatically reduces its magnitude as $\tilde{\omega}$ moves away from zero, and thus it reduces damping, i.e.~keeps $R_{\rm damp}$ closer to unity. It also means that a weaker coupling $\bar{\alpha}$ implies less damping, at least for $|\tilde{\omega}|$ not very small.

There is another effect that reduces damping for very small $|\tilde{\omega}|$, when the wavelength becomes larger than the size of the matter region. A nearly trapped wave in the cavity is a superposition of the quasi-normal modes, and these QNMs are very close to being standing waves. Their overall amplitude only slowly decreases in time due to an outgoing wave of small amplitude on the outside. Standing waves have zero net energy flux across any point $\tilde{x}$ and the energy density can vary with $\tilde{x}$ and vanish at the nodes of the wave. If $\psi(\tilde{\omega},\tilde{x})$ is the wave amplitude for the standing wave then the energy density profile $e(\tilde{\omega},\tilde{x})$ is proportional to $|\psi|^2$. We define the profile to be dimensionless with $\int_0^L e(\tilde{\omega},\tilde{x})d\tilde{x}=L$. For a 2-2-hole, waves of any angular momentum behave as S-waves at the origin, and for our radial wave equation in tortoise coordinates this implies a Dirichlet boundary condition, $\psi(\tilde{\omega},0)=0$ [1]. Thus we take the wave energy profile to be
\begin{align}
e(\tilde{\omega},\tilde{x})=2\sin(\tilde{\omega}\tilde{x})^2
.\end{align}
This simple form neglects the effect of the potential interior to the 2-2-hole, which can be small as in Fig.~\ref{fig00}; for our purposes it is sufficient that $e(\tilde{\omega},\tilde{x})$ captures the boundary condition and the essential $\tilde{\omega}$ dependence. The resonance frequencies form a discrete set $\tilde{\omega}=\tilde{\omega}_n$. For the long-lived modes the effective boundary condition at the potential barrier, $\tilde{x}=L$, is close to being Neumann and so for these modes the real part of their frequencies satisfy $\tilde{\omega}_n\approx\frac{\pi}{L}(\frac{1}{2}+n)$ for integer $n$.

\begin{figure}[h]
\centering
\includegraphics[width=0.65\textwidth]{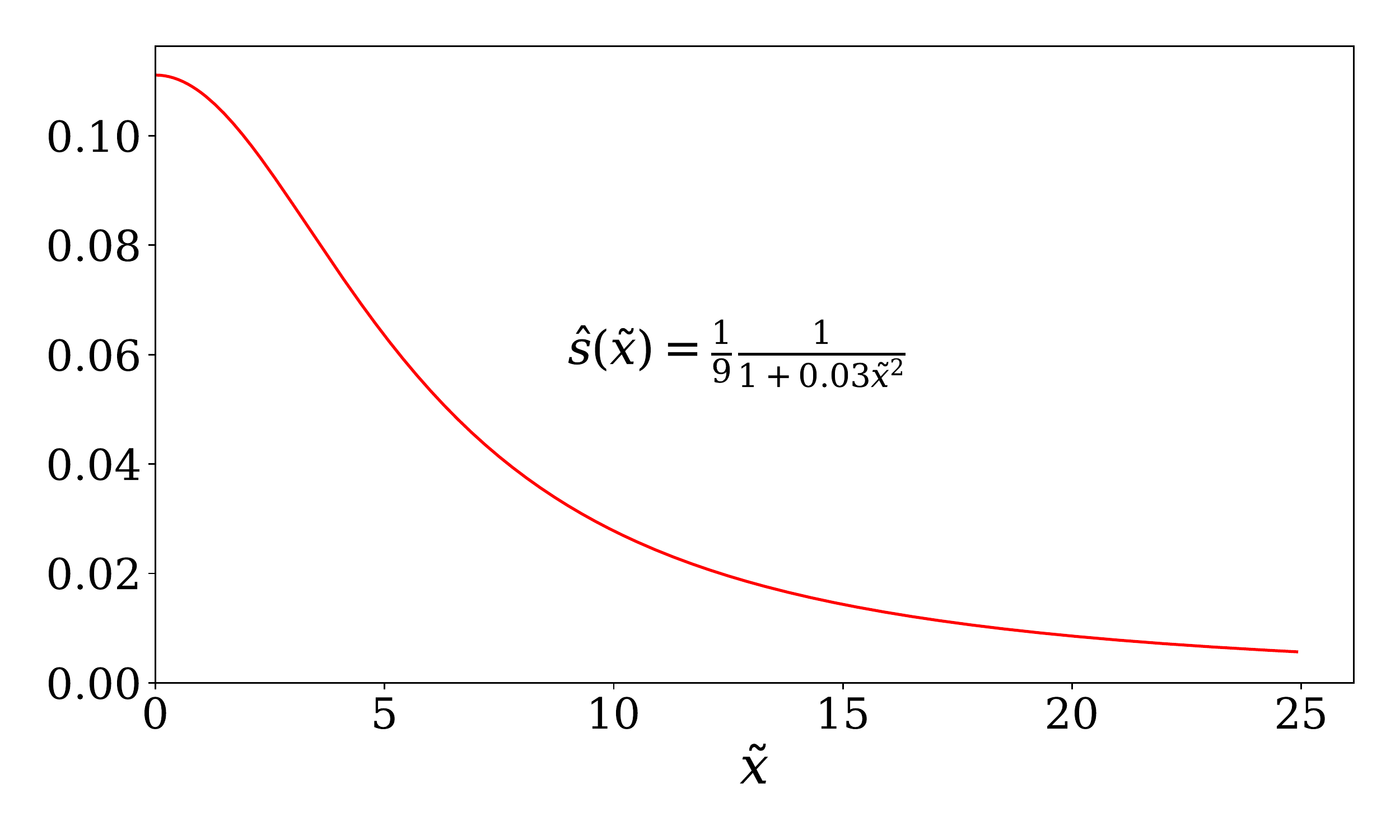}  
\vspace{-3ex}\caption{The entropy profile.}
\label{fig0}
\end{figure}

The amount of wave energy being lost locally will vary from point to point, and this will be proportional to the product of the local densities of the wave energy and the matter entropy. We may define an entropy profile $\hat{s}(\tilde{x})\propto\tilde{s}(\tilde{x})$ such that $\int_0^L \hat{s}(\tilde{x})d\tilde{x}=1$. The total damping rate thus has another $\tilde{\omega}$ dependent factor given by the overlap between these two profiles $\int_0^L \hat{s}(\tilde{x})e(\tilde{\omega},\tilde{x})d\tilde{x}$. The overlap integral is unity if $e(\tilde{\omega},\tilde{x})=1$. A numerical example of the entropy density as a function of $r$, $s(r)$, was shown in Fig.~18 of \cite{Holdom:2019ouz}. When $r$ is used instead of $x$, most of the cavity size is compressed into a small range of $r$ where $A(r)/B(r)$ spikes to a large finite value just slightly outside $2GM$. This spike is the source of the dominant log$(M)$ enhanced contribution to the cavity size $L$. Fig.~18 of \cite{Holdom:2019ouz} shows that most of the entropy is located at smaller $r$. The interior contribution to $L$ that is independent of log$(M)$ is $2\sqrt{a_2/b_2}\approx17.2$ and so Fig.~18 of \cite{Holdom:2019ouz} shows that $\hat{s}(17.2)\approx0.1\hat{s}(0)$. From this we construct an approximate estimate of the entropy profile $\hat{s}(\tilde{x})$ as shown in Fig.~\ref{fig0}. For a cavity size of $L=400$, Fig.~\ref{fig0} implies that the entropy density is concentrated near the $\tilde{x}=0$ end of the cavity as expected.

\begin{figure}[h]
\centering
\includegraphics[width=0.65\textwidth]{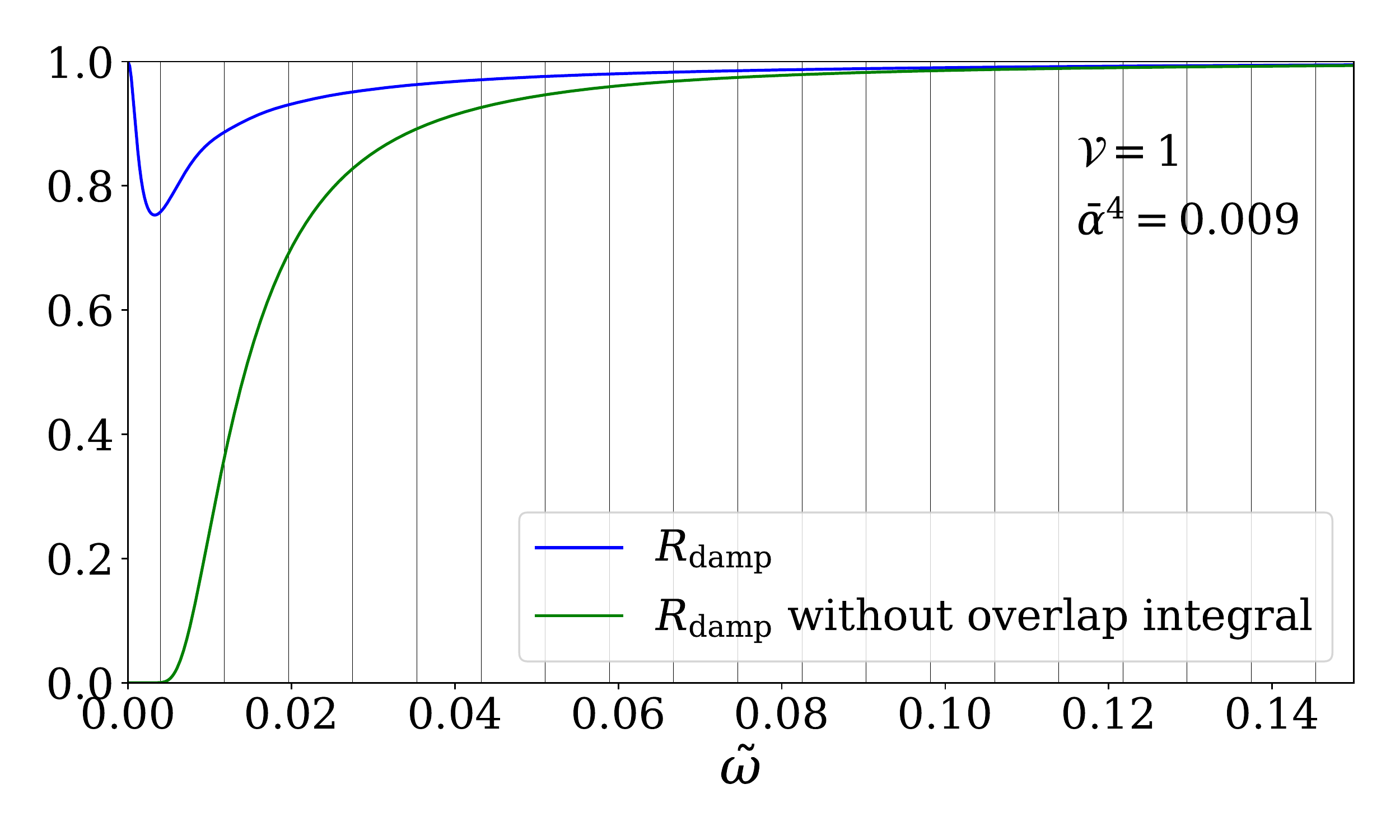}  
\caption{$R_{\rm damp}(\tilde{\omega})$ with and without the overlap integral appearing in (\ref{e6}). The vertical lines show the resonance frequencies $\frac{\pi}{L}(\frac{1}{2}+n)$ for $L=400$.}
\label{fig0a}
\end{figure}

Including the overlap integral gives our final result for the damping factor for non-rotating 2-2-holes,
\begin{align}
R_{\rm damp}(\tilde{\omega})=\exp\left(-8\pi{\cal V}(\bar{\alpha})\zeta\left[1+\left(8\pi\tilde{\omega}\frac{\zeta}{\bar{\alpha}^2}\right)^2\right]^{-1}\int_0^L \hat{s}(\tilde{x})e(\tilde{\omega},\tilde{x})d\tilde{x}\right)
.\label{e6}\end{align}
It depends on properties of the relativistic gas, the viscosity-to-entropy ratio ${\cal V}(\bar{\alpha})$ and the interaction strength $\bar{\alpha}$, as well as the fundamental parameter $\zeta$ in (\ref{e9}) that depends on the number of matter degrees of freedom. We display the damping in Fig.~\ref{fig0a} for $\zeta=2.5$, ${\cal V}=1$ and $\bar{\alpha}^2=0.009$, where the latter choices are motivated in the next section. The plot also shows the role of the overlap integral in reducing the damping. The damping that remains is only significant at small frequencies and at the lowest of the resonant frequencies.

\section{Including spin}\label{s3}
Final BHs from mergers observed at LIGO-Virgo having dimensionless spins in the $\chi\gtrsim2/3$ range. While rotating 2-2-hole solutions are not known, resonance spectra for echoes have nevertheless been obtained for this case by using an approximation for rotating 2-2-holes. The approximation is a Kerr geometry that is truncated at a new boundary situated just slightly outside the horizon. We have mentioned that the wave equation for a 2-2-hole has no angular momentum barrier at the origin $r=0$. The same is true of the truncated Kerr BH where we set $\tilde{x}=0$ at the location of the new boundary. The position of this boundary relative to the would-be horizon is again controlled by the parameter $\eta$ as in (\ref{e8}) except that the dimensionless cavity size now includes spin dependence \cite{Abedi:2016hgu,Cardoso:2017njb}
\begin{align}
   L=2\,\eta\log(\frac{GM}{\ell_{\rm Pl}})\;(\frac{1+(1-\chi^2)^{-\frac{1}{2}}}{2})
.\end{align}
Increasing spin increases the cavity size and so it causes the resonances to become even more tightly spaced.

One way that spin affects our damping result is through the effect that the new background has on the gravitational waves. We note that the radial gravitational wave equation for the Kerr background can be written in Sasaki-Nakamura form \cite{Sasaki:1981sx} where the waves continue to travel as plane waves away from the potential barrier. In the cavity these plane waves only differ by having an effective frequency $\omega-\omega_0$ and period $2\pi/|\omega-\omega_0|$. $\omega_0$ is the Kerr frequency and it is proportional to $\chi$. Thus we must replace $\tilde{\omega}$ by $\tilde{\omega}-\tilde{\omega}_0$ to have our previous discussion about frequency dependence carry over to the rotating case. We make this replacement in the damping factor (\ref{e6}) and so in the following we shall use $R_{\rm damp}(\tilde{\omega}-\tilde{\omega}_0)$.

We next reconsider the wave energy profile $e(\tilde{\omega}-\tilde{\omega}_0,\tilde{x})$ in the truncated Kerr description, but as we now argue, its basic behavior remains the same. At the new boundary we can impose a condition analogous to the Dirichlet boundary condition of the non-rotating 2-2-hole. A perfectly reflecting boundary condition is imposed by requiring that the net energy flux vanish, where the latter is the difference between the ingoing and outgoing fluxes close to the boundary. Then the wave amplitude takes the form
\begin{align}
\psi(\tilde{\omega},\tilde{x})\propto\exp(-i(\tilde{\omega}-\tilde{\omega}_0)\tilde{x})-R(\tilde{\omega})\exp(i(\tilde{\omega}-\tilde{\omega}_0)\tilde{x})
,\end{align}
where $R(\tilde{\omega})$ is a known non-negative real function \cite{Conklin:2017lwb}. From this we determine the energy density profile,
\begin{align}
e(\tilde{\omega},\tilde{x})=\frac{4R(\tilde{\omega})\sin((\tilde{\omega}-\tilde{\omega}_0)\tilde{x})^2+(R(\tilde{\omega})-1)^2}{1+R(\tilde{\omega})^2}
.\end{align}
The key point here is that $R(\tilde{\omega}_0)=1$, and as well $R(\tilde{\omega})$ is not rapidly varying. Thus for frequencies close to $\tilde{\omega}_0$ (where the wavelength is large) the energy profile is still small where the entropy profile is large. And as $\tilde{\omega}\to\tilde{\omega}_0$ the overlap integral still tends to zero. In fact the overlap integral has little impact unless $\tilde{\omega}$ is close to $\tilde{\omega}_0$, and so there is very little dependence on the behavior of $R(\tilde{\omega})$ away from $\tilde{\omega}_0$.

Further spin dependence can enter the damping factor through the thermodynamics of the relativistic gas in a rotating 2-2-hole. In particular we need the quantities $S/{\rm Area}=\zeta_1(\chi)/4G$ and $T_\infty=T^{\rm Kerr}_{\rm Hawking}(\chi)/\zeta_2(\chi)$. We define $T_\infty$ with respect to the known spin dependence of the Hawking temperature of the Kerr BH, $T^{\rm Kerr}_{\rm Hawking}(\chi)=2T_{\rm Hawking}\sqrt{1-\chi^2}/(1+\sqrt{1-\chi^2})$. In the following we shall present results with $\zeta_1(\chi)=\zeta_2(\chi)=\zeta$, and we comment more on the neglect of this unknown spin dependence in the Conclusions.

\begin{figure}[h]
   \centering
      \includegraphics[width=0.48\textwidth]{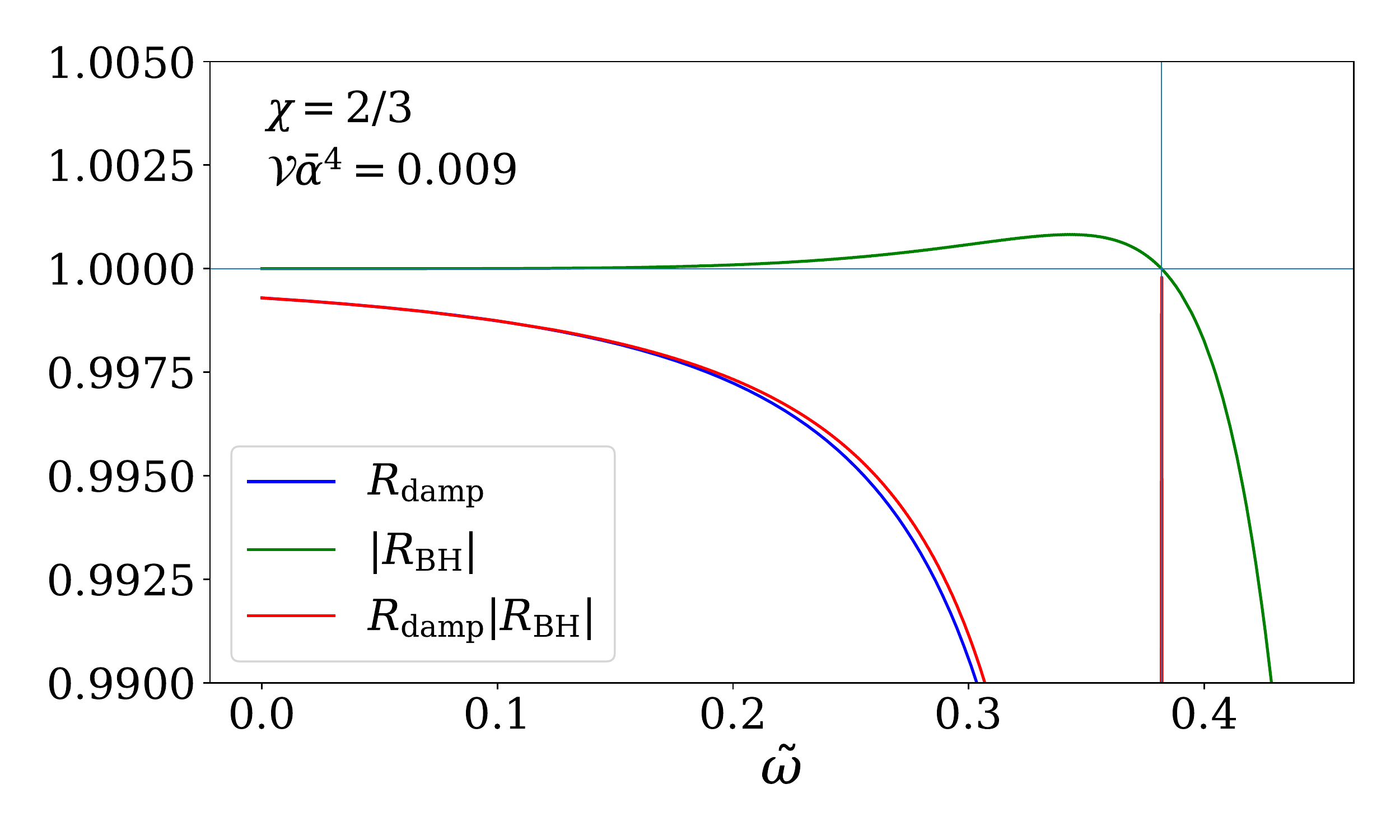} 
      \includegraphics[width=0.48\textwidth]{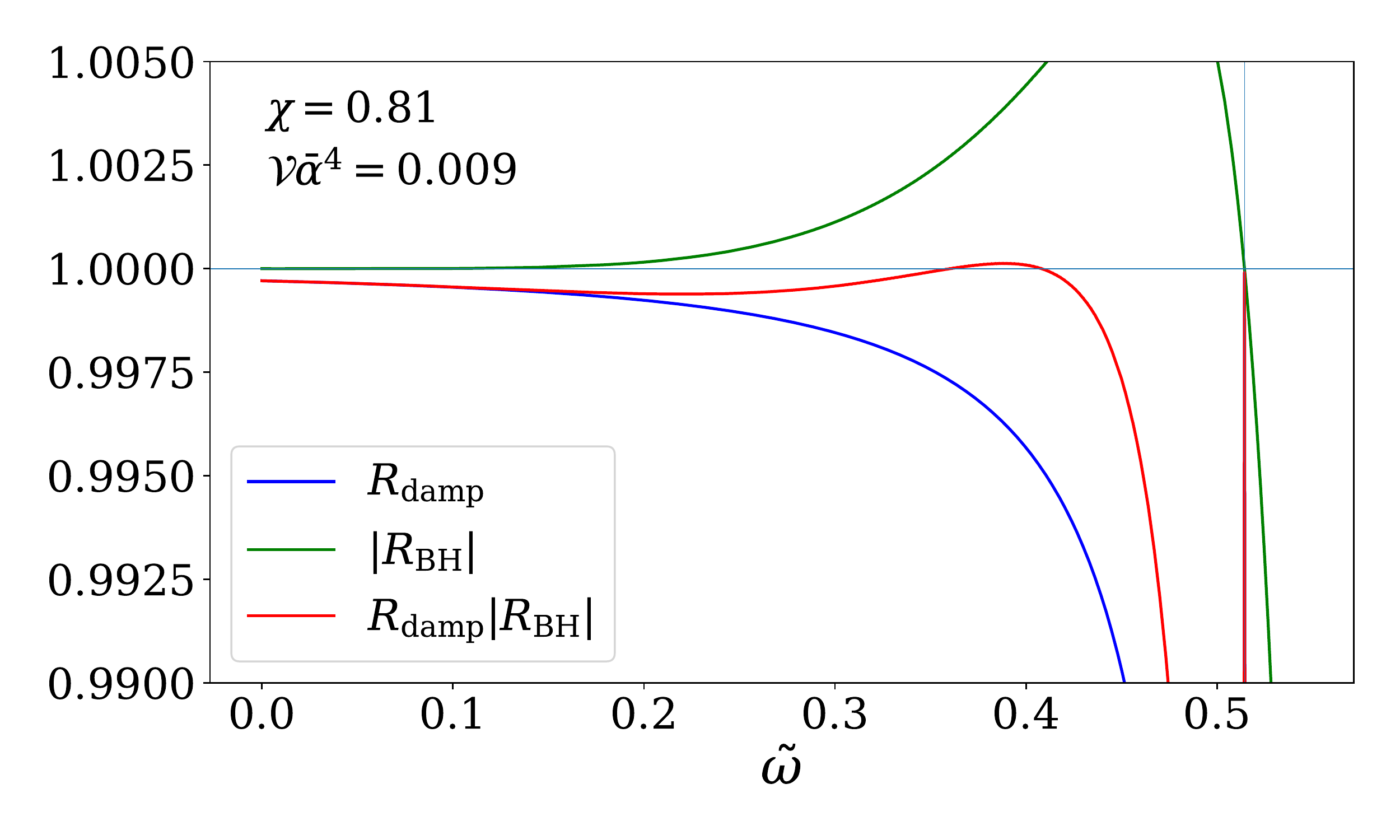}  
\caption{The damping factor $R_{\rm damp}$ is compared to the Kerr BH reflection amplitude $|R_{\rm BH}|$ and their product. The frequency where $R_{\rm damp}$ spikes back up is $\tilde{\omega}_0$. The two plots compare results for different spins $\chi=2/3$ and $\chi=0.81$.}
   \label{fig1}
   \end{figure}
   
In Fig.~\ref{fig1} we display $R_{\rm damp}(\tilde{\omega}-\tilde{\omega}_0)$ along with $R_{\rm BH}(\tilde{\omega})$, the reflection amplitude for a Kerr BH. $|R_{\rm BH}|>1$ implies superradiant amplification and a possible ergoregion instability for the horizonless 2-2-hole. The product $|R_{\rm BH}|R_{\rm damp}$ is also shown in the figure and when $|R_{\rm BH}|R_{\rm damp}<1$, the damping is sufficient to remove the instability. For the modeling of the spectra in \cite{Holdom:2019bdv} we used the equivalent of a constant $R_{\rm damp}$ (more precisely a partially absorbing wall as described by $|R_{\rm wall}|=0.995$ and 0.992 for $\chi=2/3$ and 0.81 respectively). Our new more realistic $R_{\rm damp}(\tilde{\omega}-\tilde{\omega}_0)$ quite quickly approaches unity for frequencies away from $\tilde{\omega}_0$. The $R_{\rm damp}(\tilde{\omega}-\tilde{\omega}_0)$ dependent curves that are visible in Fig.~\ref{fig1} are essentially only dependent on the combination ${\cal V}(\bar{\alpha})\bar{\alpha}^4$. When this has a value of 0.009 the left plot with $\chi=2/3$ shows that the instability is easily removed. In the right plot with higher spin $\chi=0.81$ we see that the same 0.009 value is now on the stability edge. Given that event GW170729 has such a spin, this sets a bound ${\cal V}(\bar{\alpha})\bar{\alpha}^4\gtrsim0.009$.

\begin{figure}[h]
   \centering
      \includegraphics[width=0.48\textwidth]{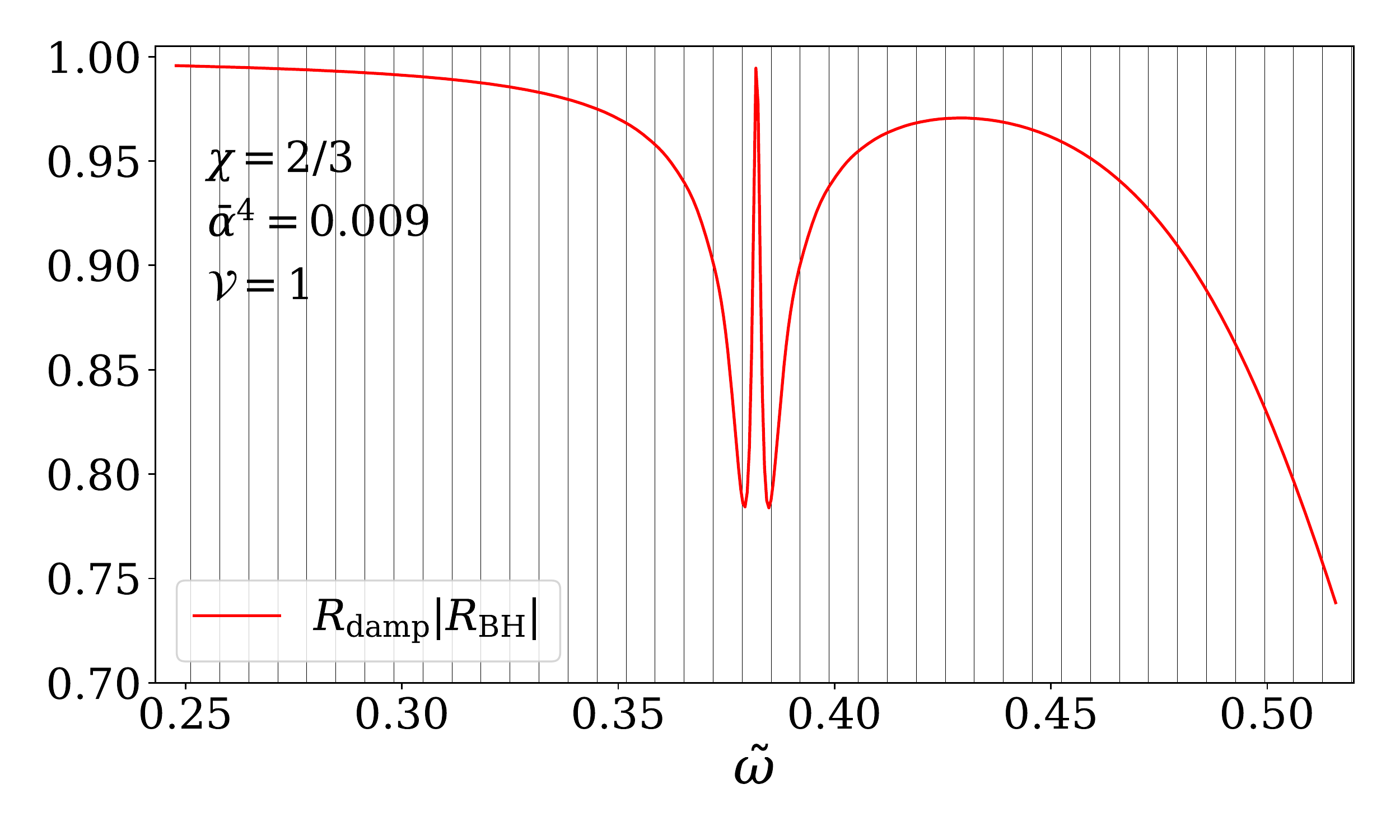} 
      \includegraphics[width=0.48\textwidth]{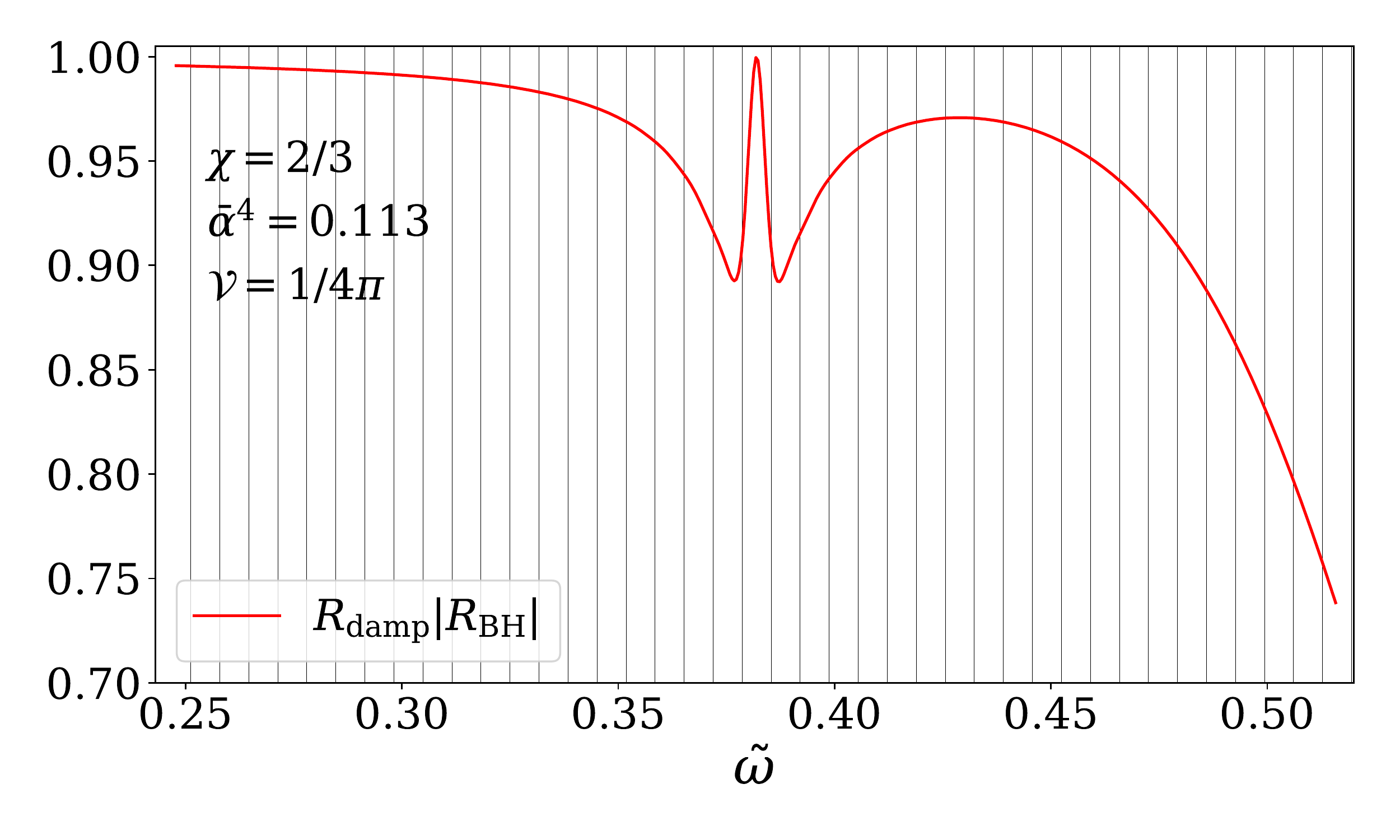}  
\caption{For $\chi=2/3$, the vertical and horizontal plot ranges are changed from Fig.~\ref{fig1}. The two plots show different choices of ${\cal V}$ and $\bar{\alpha}$ with ${\cal V}\bar{\alpha}^4=0.009$. The vertical lines show the resonance frequencies $\tilde{\omega}_0+\frac{\pi}{L}(\frac{1}{2}+n)$.}
\label{fig2}
\end{figure}

In Fig.~\ref{fig2} we focus on the behavior of $R_{\rm damp}|R_{\rm BH}|$ around $\tilde{\omega}_0$. The two plots show this product for two choices of $\bar{\alpha}$ for fixed ${\cal V}(\bar{\alpha})\bar{\alpha}^4$. The plot on the right has ${\cal V}(\bar{\alpha})=1/(4\pi)$, the minimum value, and a correspondingly larger $\bar{\alpha}$, and we see that this produces the minimal amount of damping for fixed ${\cal V}(\bar{\alpha})\bar{\alpha}^4$.

\begin{figure}[h]
   \centering
      \includegraphics[width=0.48\textwidth]{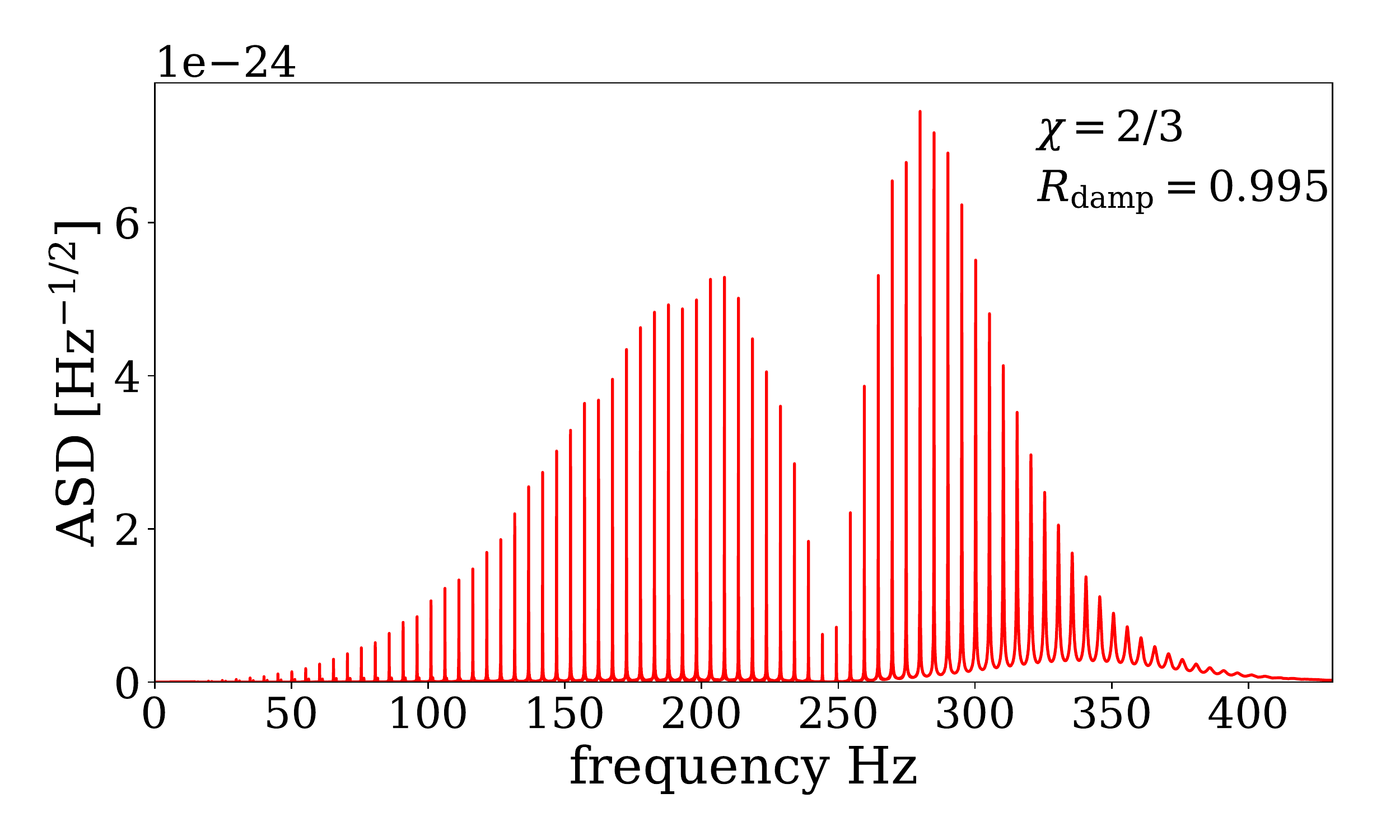} 
      \includegraphics[width=0.48\textwidth]{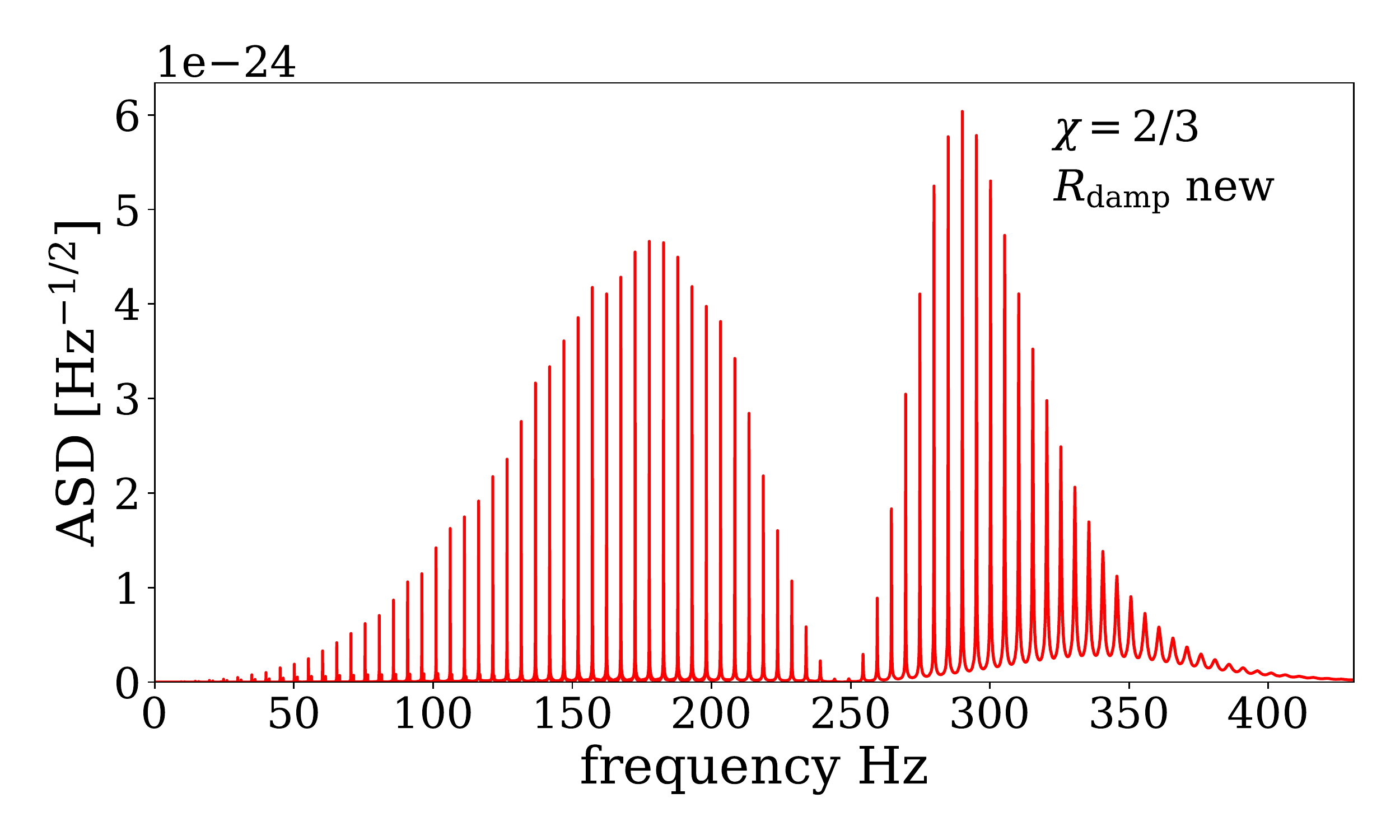}  
\caption{The spectra for a spin of $\chi=2/3$ with 180 echoes are shown as amplitude spectral densities. The left plot uses a constant $R_{\rm damp}$ and the right plot uses our new $R_{\rm damp}(\tilde{\omega}-\tilde{\omega}_0)$ when ${\cal V}=1$ and $\bar{\alpha}^4=0.009$. We have converted frequencies to Hz by using a (detector frame) mass of $50 M_\odot$. The spectra are normalized so that the peak strain in the time domain is $5\times10^{-22}$.}
\label{fig3}
\end{figure}

In Fig.~\ref{fig3} we show the effect that the new $R_{\rm damp}(\tilde{\omega}-\tilde{\omega}_0)$ has on the observable spectrum. This is the spectrum reconstructed from the real part of the strain ${\rm Re}(h(t))$ over some time duration, corresponding to 180 echoes in this example. A Greens function approach in frequency space produces a spectrum that is a product of a transfer function and a source integral, and it is from this that $h(t)$ is obtained \cite{Conklin:2019fcs,Holdom:2019bdv}. The source integral is controlled by the spectral distribution of the initial perturbation, and for our example here we take a distribution proportional to $\exp(-20(\tilde{\omega}-\tilde{\omega}_0)^2)$ for a perturbation that arises inside the cavity. The source integral also brings in an explicit factor of $\omega-\omega_0$ that suppresses the spectrum around $\omega_0$. The two plots compare the use of our previously used constant $R_{\rm damp}$ to our new $R_{\rm damp}(\tilde{\omega}-\tilde{\omega}_0)$. We see how the latter produces increased suppression around $\tilde{\omega}_0$ and less suppression elsewhere. The $\bar{\alpha}$ dependence for fixed ${\cal V}(\bar{\alpha})\bar{\alpha}^4$ illustrated in the Fig.~\ref{fig2} has very little effect since it is occurring in an already suppressed region. Thus the observed spectra essentially just constrains ${\cal V}(\bar{\alpha})\bar{\alpha}^4$.

\section{Conclusions}
A relativistic gas is the matter source for a 2-2-hole, and this displays a viscosity that can cause damping of gravitational waves. But the collisional frequency in the gas is typically less than the wave frequency and this drastically reduces the damping for those wave frequencies. For large wavelengths there is another reduction since the wave amplitude is small where the matter is located. The end result is that the remaining damping is largest for relatively low frequencies. When spin is introduced then the damping is largest for similarly low $|\omega-\omega_0|$. The superradiant effect occurs for $\omega<\omega_0$ at a $|\omega-\omega_0|$ that is just somewhat larger, and so the damping can play a role in taming the ergoregion instability. We also displayed the effect of damping on the echo resonance spectrum; the effect is not large since the spectrum is already modulated by a $|\omega-\omega_0|$ factor. Overall, the small size of the damping strengthens the case for echo searches that consider many echoes, which is when the resonance spectrum becomes very striking as in Fig.~\ref{fig3}.

We have identified a quantity ${\cal V}(\bar{\alpha})\bar{\alpha}^4$ that controls the size of the damping relevant for the ergoregion instability, where ${\cal V}(\bar{\alpha})$ is the viscosity-to-entropy ratio and $\bar\alpha$ is a coupling that characterizes the particle interactions in the gas. If we re-introduce the unknown spin dependence as mentioned in the last section, then this becomes ${\cal V}(\bar{\alpha})\bar{\alpha}^4\zeta_1(\chi)\zeta/\zeta_2(\chi)^2$. Depending on whether this quantity increases or decreases with spin, the effect would be to weaken or strengthen the instability at high spin.

There may still be some critical value of the spin above which the ergoregion instability is not completely damped. It could be that the instability extracts energy from the rotational energy and that an 2-2-hole with a high spin will spin down to near the critical spin. The interesting question is how violent this process is. Another question is the effect of accretion on the instability. Accretion means that matter at a temperature much higher than $T_\infty$ is entering the cavity. Until this matter equilibrates with the matter already in the cavity, it can cause gravitational wave damping unsuppressed by either of the two suppression mechanisms we have identified. It remains to determine the extent to which this can increase stability at high spins.

Our analysis potentially constrains ${\cal V}(\bar{\alpha})\bar{\alpha}^4$ both from above and from below. Too small and the smaller damping may lead to an instability for observed high spins, e.g.~for GW170729. Too large and it causes quite pronounced damping in a large region around $\omega_0$, and this can potentially be constrained by observations of echoes. This type of restriction on ${\cal V}(\bar{\alpha})\bar{\alpha}^4$ is interesting since it characterizes the fundamental interactions among particles at Planck-like temperatures and beyond. More theoretical knowledge of the coupling dependence of ${\cal V}(\bar{\alpha})$ could then separately constrain both this quantity and the value of the effective coupling.

\begin{acknowledgements}
   This research was supported in part by the Natural Sciences and Engineering Research Council of Canada.
 \end{acknowledgements}


\begin{thebibliography}{99}
\bibitem{Holdom:2016nek} 
  B.~Holdom and J.~Ren,
  ``Not quite a black hole'',
  Phys.\ Rev.\ D {\bf 95}, no. 8, 084034 (2017).
  [arXiv:1612.04889 [gr-qc]].

 \bibitem{Holdom:2019ouz} 
  B.~Holdom,
  ``A ghost and a naked singularity; facing our demons,''
  arXiv:1905.08849 [gr-qc].
  
\bibitem{Ren:2019jft} 
  J.~Ren,
  ``Anatomy of a thermal black hole mimicker,''
  Phys.\ Rev.\ D {\bf 100}, no. 12, 124012 (2019)
  [arXiv:1905.09973 [gr-qc]].

  \bibitem{Conklin:2017lwb} 
  R.~S.~Conklin, B.~Holdom and J.~Ren,
  ``Gravitational wave echoes through new windows'',
  Phys.\ Rev.\ D {\bf 98}, no. 4, 044021 (2018)
  [arXiv:1712.06517 [gr-qc]].

\bibitem{Cardoso:2016rao} 
  V.~Cardoso, E.~Franzin and P.~Pani,
  ``Is the gravitational-wave ringdown a probe of the event horizon?,''
  Phys.\ Rev.\ Lett.\  {\bf 116}, no. 17, 171101 (2016)
  Erratum: [Phys.\ Rev.\ Lett.\  {\bf 117}, no. 8, 089902 (2016)]
  [arXiv:1602.07309 [gr-qc]].

\bibitem{Cardoso:2016oxy} 
V.~Cardoso, S.~Hopper, C.~F.~B.~Macedo, C.~Palenzuela and P.~Pani,
``Gravitational-wave signatures of exotic compact objects and of quantum corrections at the horizon scale'',
Phys.\ Rev.\ D {\bf 94}, no. 8, 084031 (2016)
[arXiv:1608.08637 [gr-qc]].

\bibitem{Abedi:2016hgu} 
  J.~Abedi, H.~Dykaar and N.~Afshordi,
  ``Echoes from the Abyss: Tentative evidence for Planck-scale structure at black hole horizons'',
  Phys.\ Rev.\ D {\bf 96}, no. 8, 082004 (2017)
  [arXiv:1612.00266 [gr-qc]].

\bibitem{Holdom:2019bdv}
   B.~Holdom,
   ``Not quite black holes at LIGO,''
   Phys.\ Rev.\ D \textbf{101}, no.6, 064063 (2020)
   [arXiv:1909.11801 [gr-qc]].

\bibitem{fried}
J.~L.~Friedman, “Ergosphere instability,” Communications in Mathematical Physics 63 (Oct, 1978) 243–255.
   
\bibitem{Brito:2015oca}
R.~Brito, V.~Cardoso and P.~Pani,
``Superradiance: New Frontiers in Black Hole
Physics,''
Lect. Notes Phys. \textbf{906}, pp.1-237 (2015)
[arXiv:1501.06570 [gr-qc]].

\bibitem{Maggio:2017ivp}
E.~Maggio, P.~Pani and V.~Ferrari,
``Exotic Compact Objects and How to Quench their Ergoregion Instability,''
Phys. Rev. D \textbf{96} (2017) no.10, 104047
[arXiv:1703.03696 [gr-qc]].

\bibitem{Maggio:2018ivz}
E.~Maggio, V.~Cardoso, S.~R.~Dolan and P.~Pani,
``Ergoregion instability of exotic compact objects: electromagnetic and gravitational perturbations and the role of absorption,''
Phys. Rev. D \textbf{99}, no.6, 064007 (2019)
[arXiv:1807.08840 [gr-qc]].

\bibitem{Hawking:1966qi}
S.~Hawking,
``Perturbations of an expanding universe,''
Astrophys.\ J.\  \textbf{145} 544 (1966).

\bibitem{Arnold:2000dr}
P.~B.~Arnold, G.~D.~Moore and L.~G.~Yaffe,
``Transport coefficients in high temperature gauge theories. 1. Leading log results,''
JHEP \textbf{11}, 001 (2000)
[arXiv:hep-ph/0010177 [hep-ph]].

  \bibitem{Conklin:2019fcs} 
  R.~S.~Conklin and B.~Holdom,
 ``Gravitational wave echo spectra,''
  Phys.\ Rev.\ D {\bf 100}, no. 12, 124030 (2019)
  [arXiv:1905.09370 [gr-qc]] (and references therein).

\bibitem{Cardoso:2017njb} 
  V.~Cardoso and P.~Pani,
  ``The observational evidence for horizons: from echoes to precision gravitational-wave physics'', Nat.\ Astron.\ 1, 586 (2017)
  [arXiv:1707.03021 [gr-qc]].

\bibitem{Kovtun:2004de} 
  P.~Kovtun, D.~T.~Son and A.~O.~Starinets,
  ``Viscosity in strongly interacting quantum field theories from black hole physics,''
  Phys.\ Rev.\ Lett.\  {\bf 94}, 111601 (2005)
  [hep-th/0405231].

\bibitem{Loeb:2020lwa} 
  A.~Loeb,
  ``Upper Limit on the Dissipation of Gravitational Waves in Gravitationally Bound Systems,''
  Astrophys.\ J.\  {\bf 890}, no. 2, L16 (2020)
  [arXiv:2001.01730 [astro-ph.HE]].

\bibitem{Sasaki:1981sx}
M.~Sasaki and T.~Nakamura,
``Gravitational Radiation From a Kerr Black Hole. 1. Formulation and a Method for Numerical Analysis,''
Prog.\ Theor.\ Phys.\  {\bf 67}, 1788 (1982).


\end{thebibliography}
\end{document}